# Superradiance Enhanced Light-Matter Interaction in Spatially Ordered Shape and Volume Controlled Single Quantum Dots: Enabling On-Chip Photonic Networks

*Lucas Jordao, Swarnabha Chattaraj[†], Qi Huang, Siyuan Lu[‡], Jiefei Zhang[†] & Anupam Madhukar\**

Nanostructure Materials and Devices Laboratory, University of Southern California, Los Angeles, CA 90089-0241, United States of America

[†] Currently at Argonne National Laboratory. Illinois. 60439, United States of America.

[‡] Currently at Xmotors.ai. California. 95054, United States of America.

Corresponding Author Email: madhukar@usc.edu

**ABSTRACT.** On-chip photonic networks require adequately spatially ordered matter-photon interconversion qubit sources with emission figures-of-merit exceeding the requirements that would enable the desired functional response of the network. The mesa-top single quantum dots (MTSQDs) have recently been demonstrated to meet these requirements. The substrate-encoded size-reducing epitaxy (SESRE) approach underpinning the realization of these unique quantum emitters allows control on the shape, size, and strain (lattice-matched or mismatched) of these epitaxial single quantum dots. We have exploited this unique feature of the MTSQDs to reproducibly create arrays of quantum dots that exhibit single photon superradiance, a characteristic of the SESRE-enabled delicate balance between the confinement potential volume, depth, the resulting exciton binding energy, and the degree of confinement of the center of mass (CM) motion of the exciton. Scanning transmission electron microscope (STEM) studies reveal the structural (atomic scale) and chemical (nm scale) nature of the material region defining the notion of the shape and volume (here large) of the electron confinement region (*i.e.* the QD). In the exciton's weak CM confinement regime, owing to its coherent sampling of the large volume, enhancement of the MTSQD oscillator strength to ~30 is demonstrated. Theoretical modelling with input from the STEM findings provide corroboration for single photon superradiance causing enhancement of the oscillator strength by ~2.5 to 3. Our findings allow fabricating and studying interconnected networks enabled by these unique matter qubit-light qubit interconversion units that can be realized for lattice matched and mismatched material combinations covering UV to midinfrared wavelength range.

## 1 INTRODUCTION

Conversion of information from matter qubit to light qubit back and forth with the highest fidelity is at the core of all quantum information processing (QIP) hardware approaches [1–3]. This conversion, in turn, is controlled by light-matter interaction. Thus, implementation systems and approaches that can tailor light-matter coupling are of considerable significance to QIP systems employing any of the major exploited physical hardware platforms: matter qubits represented in atoms, ions, structural and/or chemical defects in solids, semiconductor quantum dots, and Josephson junction based superconducting circuits. The conceptual and operational physics of matter-photon qubit conversion in these platforms has usually been modelled as an electric dipole driven transition in an effective two-level matter system. It has guided the interpretation of the transition rate ($T_1^{-1}$) as the product of the transition oscillator strength, $f(\omega)$, and the available local density of photon states at the transition frequency ($\omega$), $\rho(r_0, \omega)$ [4].

$$T_1^{-1} \propto f(\omega) * \rho(r_0, \omega) \qquad (1)$$

For most matter qubits, such as created in atoms, ions and solid-state emitters (defects / deep levels) the two-level transition oscillator strength $f$ is essentially fixed [5–7] and the transition (decay) rate is manipulated primarily by tailoring the local photon density of states, $\rho(r_0, \omega)$. This is achieved through modification of the dielectric environment around the emitter by such means as embedding the emitter in a cavity and / or waveguide designed to *enhance* the local density of photon states to which photons couple, thereby enhancing the matter qubit (typically exciton) decay rate. By contrast, semiconductor quantum dots (QDs) are a unique class of quantum emitters in which the oscillator strength for exciton decay itself can be manipulated (enhanced) through control on the relative volume and strength (depth) of the confinement potential and the resulting binding energy and volume of the exciton formed by the excitation of the electron from the confined highest valence band derived state to the lowest confined conduction band derived electron state [8]. This is because of the *single photon* superradiance effect [9, 10] which, amongst the inorganic quantum emitters under investigation for quantum information, is realizable only in QDs as these can be tailored to exhibit weak confinement of the exciton's center-of-mass in a volume larger than its own, leading to enhancement of the atomic oscillator strength arising from energy storage in a *coherent* collective quantum state shared across the atoms of the confining volume of the dot. The enhanced oscillator strength, $f$, in turn gives enhanced light-matter interaction.

Strong light-matter interaction is particularly important to developing multiple emitter-based quantum networks as network system-level performance imposes strict requirements on the characteristics of the individual quantum emitters constituting the platform to be employed [11]. As discussed in ref.11, the individual emitter's single photon characteristics must consist of near unity quantum efficiency, single photon purity, and indistinguishability in order to meet the requirements for QIP applications such as linear optical quantum computing (LOQC) and Boson sampling [12, 13]. Beyond these individual characteristics, interconnecting emitters to realize system-level quantum circuits/networks for QIP applications demands: (1) designed on-chip positioning of the emitters to nm accuracy for optical wavelength regime, and (2) emission wavelength nonuniformity of the emitters within the threshold allowed for on-chip tuning technologies (*e.g.* ~1V to 3V applied bias for a ~3nm wavelength shift via the Stark effect) [14, 15] thereby enabling multi-photon interference. Strong light-matter interaction, albeit not a strict requirement for such platforms, is highly beneficial as the resultant faster radiative decay lifetime of the emitters would lead to increased robustness to intrinsic dephasing [16]. This would also allow the system to operate at higher frequencies [17, 18].

To date the main limitation in achieving aforementioned platform with quantum dot-based emitters has come from the lack of *adequate* control over the QD positions and their size, shape and composition (*i.e.* the effective 3D confinement potential) across the grown sample due to the random nature of the process by which the most popular employed epitaxial QDs are synthesized -- the lattice-mismatched strain-driven self-assembled quantum dots (SAQDs) [19, 20] and the droplet epitaxy quantum dots (DEQDs) [21, 22]. The lack of adequate spatial positioning precludes developing on-chip quantum optical circuits and the lack of adequate control on size, shape, and volume prevents exploiting the benefits of enhanced light-matter interaction arising from superradiance made possible by the mesoscopic nature of QDs.

We present here scanning transmission electron microscopy (STEM) and energy dispersive x-ray spectroscopy (EDS) based structural / compositional findings on a unique class of QDs, the mesa top single quantum dots (MTSQDs), that are not only synthesized in adequately accurate controlled locations, with spectral emission characteristics which satisfy all individual- and system-level requirements for QIP [11], *but that also* show reproducible, and controllable, large oscillator strengths resulting from single-photon superradiance. Our results on the control over MTSQD positioning, size, shape, volume, and thus the resulting confinement potential depth and profile (across the typically ill-defined heterojunction surface), enabled by the *substrate encoded size reducing epitaxy* (SESRE) [23–25] growth approach employed, allow for reproducibly synthesizing scalable arrays of quantum emitters with enhanced light-matter interaction. Indeed, neutral exciton radiative decay lifetimes $T_1 < 400$ps and large oscillator strengths ($f \sim 30$) are demonstrated for large arrays. Moreover, MTSQD's fast radiative decay rates allow for remarkable robustness to intrinsic phonon-dephasing which has led to the demonstration of high single photon indistinguishability and high 2-qubit CNOT gate operation fidelity [26]. These results underpin MTSQD's high promise as technologically relevant platform for QIP.

## 2 RESULTS AND DISCUSSION

### 2.1 Spatially Ordered Arrays of Large-Volume Shape-Controlled Superradiant MTSQDs

The SESRE approach based MTSDQs in arrays were examined for their structure and composition using STEM and EDS at various resolutions, reaching atomic. Results on the structural characterization of large-volume shape-controlled MTSQDs are shown in Fig. 1 (details on the sample structure and growth are given in the Experimental Methods section). A low-magnification STEM image of the TEM lamella specimen prepared from a row of MTSQDs in a 5x8 array is shown in Fig. 1a, and Z-contrast STEM images of the individual nanomesas, from the lamella specimen of Fig. 1a, are shown in Figs. 1b – d. The growth front profile evolution of SESRE growth on the pedestal nanomesas is illustrated in Figs. 1b – d, where the dark lines are the AlAs marker layers and the grey contrast is GaAs. As material is deposited on the nanomesas, {103} sidewall facets form and the (001) mesa top size lateral reduces owing to the surface stress gradient-driven preferential migration of cations (Ga and Al) from the sidewall to the top and incorporation of adatoms on the [001] mesa top. Once the mesa top size reduces to ~30nm, $In_{0.5}Ga_{0.5}As$ is deposited to form the MTSQD as revealed by the white contrast at the apex of the nanomesas on the STEM images of Figs. 1b – d. The position accuracy of each MTSQD with respect to the nanomesa's center is ~3nm laterally (in the [110] direction) and ~1nm vertically (in the [001] direction). This high position accuracy of MTSQDs is due to the control on SESRE growth discussed below where all nanomesas evolve with the same profile during growth resulting in MTSQDs always forming centered at the apex of the nanomesas. The overall positioning accuracy of MTSQDs with respect to each other within the array is likely a bit higher and is limited by the spatial resolution of EBL (~5nm) patterning and fluctuations in size of the *as-etched* nanomesas in the array.

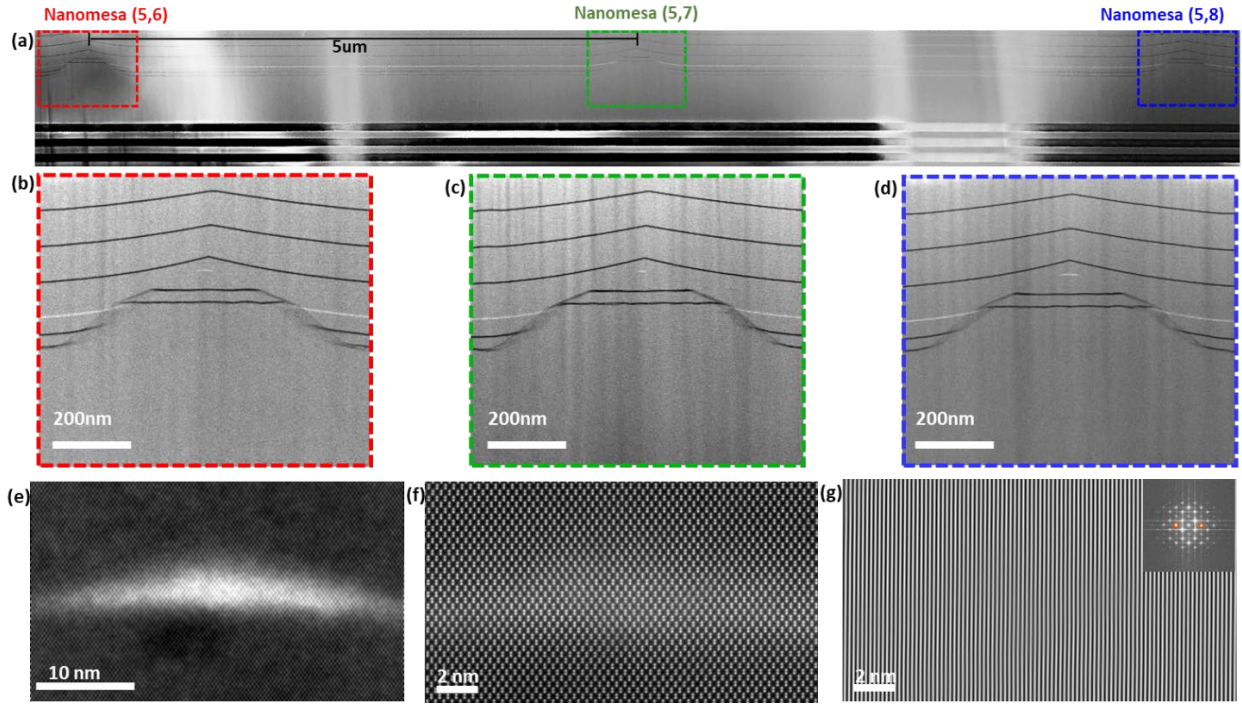

**Figure 1. Structural characterization of spatially ordered arrays of large-volume and shape-controlled MTSQDs in a 5 x 8 array.** **(a)** Low-magnification [110] cross-sectional Z-contrast STEM image (top panel) of a TEM lamella specimen with three nanomesas, containing MTSQDs, from an array. **(b), (c), (d)** Z-contrast STEM images of the nanomesas in (a) revealing the growth front evolution during SESRE growth and the MTSQDs at the apex of the nanomesas. **(e)** Z-contrast STEM image (with Gaussian Blur filter) of MTSQD (5,6) showing its large-volume and unique shape. **(f)** Atomic-resolution HAADF STEM image of defect-free MTSQD (5,6). **(g)** Fourier/Bragg filtered image of (f) using the {220} "Bragg" spots (inset shows the FFT of (f), with orange circles indicating where the mask for the filtering was applied).

MTSQDs are grown via the SESRE approach which ensures reproducible control of their size and shape across the spatially-ordered arrays. Briefly, SESRE growth of MTSQDs comprises three stages: (i) nanomesa top size reduction, (ii) QD formation at the mesa top and nanomesa pinch-off, and (iii) surface morphology planarization. All these stages of growth are captured by the STEM images in Figs. 1b – d. In stages (i) and (ii) of growth, interfacet migration of adatoms from sidewalls to mesa top and preferential incorporation of adatoms leads to mesa top size reduction, enabling deposition of QD material at a mesa top size appropriate for single QD formation and resulting in the formation of an MTSQD near the apex of the nanomesa before pinch-off. Continued growth allows the nanomesa pyramidal morphology to pinch-off [23] and be subsequently planarized [27] during stage (iii) of growth, as shown by the continuous profile of the AlAs marker layers deposited after the MTSQD. Planarization is achieved due to the nanomesa pedestal morphology, with {101} base facets, which allows for a reversal of adatom migration after nanomesa pinch-off; the shallow {100} vertical sidewalls of the nanomesa provide a contiguous link between the {103} sidewalls and the {101} base facets permitting adatoms to migrate away from the nanomesa sidewalls and towards surrounding planar regions. This adatom migration away from the nanomesa increases the growth rate of the planar region with respect to the nanomesa pyramidal morphology and thus buries the nanomesa [27].

Fig. 1e shows a Z-contrast STEM image of the MTSQD on nanomesa (5,6) revealing the MTSQDs large-volume and shape. A Gaussian blur filter was applied to the STEM image to enhance the Z-contrast between the InGaAs MTSQD and the surrounding GaAs, better showcasing the MTSQD's size and shape. The size and shape of MTSQDs formed during InGaAs deposition is strongly influenced by: (i) nanomesa's top size and morphology, and (ii) growth conditions employed. From STEM characterization it is observed that the base length of MTSQDs along [110] cross-section is roughly equal to the nanomesa opening top size when the QD material (InGaAs) is deposited. Additionally, we are able to employ reproducible growth conditions in our SESRE growths by using the machine-condition transfer-function approach [28], which ensures that growth conditions, and the resulting nanomesa profile evolution, can be controlled from run-to-run. Therefore, by controlling the nanomesa profile evolution, and thus the mesa top opening size before QD deposition, we can control the size and shape of MTSQDs formed. The large-volume and distinct shape of the MTSQD shown in Fig. 1e is observed consistently across different MTSQDs inspected in the arrays. This highlights the ability to control the MTSQD shape and size across the arrays, a feature unique to SESRE growth, and allows for tailoring light-matter interaction in the MTSQDs.

An atomic-resolution STEM image of the MTSQD in Fig. 1e is shown in Fig. 1f. The individual Ga, As and In atomic-columns, with spacing of ~0.141nm projected along the [1-10] direction (*i.e.* the electron beam direction), are clearly resolved and show that the MTSQD is defect-free. The atomic columns containing In atoms appear brighter than columns containing only Ga and As atomic columns due to In having a larger atomic number; In atoms have larger elastic scattering cross-sections than Ga and As atoms and so more signal is collected at the high-angle annular dark field (HAADF) detector when the beam scans across regions with In atoms. Furthermore, to further examine for the presence of extended defects such as dislocations or stacking faults, the STEM image from Fig. 1f was Fourier/Bragg filtered and the result is shown in Fig. 1g. The image in Fig. 1g was produced using the {220} Bragg spots and shows that the {220} lattice fringes in the MTSQD and the surrounding GaAs are without any discontinuities, evidencing that the MTSQD is defect-free, *i.e.* coherent.

The reproducible nature of MTSQD growth allows us to make inferences about and correlations to their optical behavior, such as the observed large oscillator strengths [29], as discussed next.

*2.2 Scaling of Array Size of Spectrally Uniform Shape-Controlled Large Volume MTSQDs*

Results on the spectral inhomogeneity of scalable arrays of large-volume shape-controlled MTSQDs are shown in Fig. 2. Spectrally resolved large area photoluminescence (PL) imaging of ~1400 emitters from a (50x50) array (a portion of the 2500 MTSQDs in the array, limited by the excitation area) is shown in Fig. 2a. Analysis of the wavelength resolved image in Fig. 2a shows that more than 99% of the MTSQDs are emitting and that the spectral nonuniformity in the array is low, $\sigma_\lambda$<5nm (Fig. 2b).

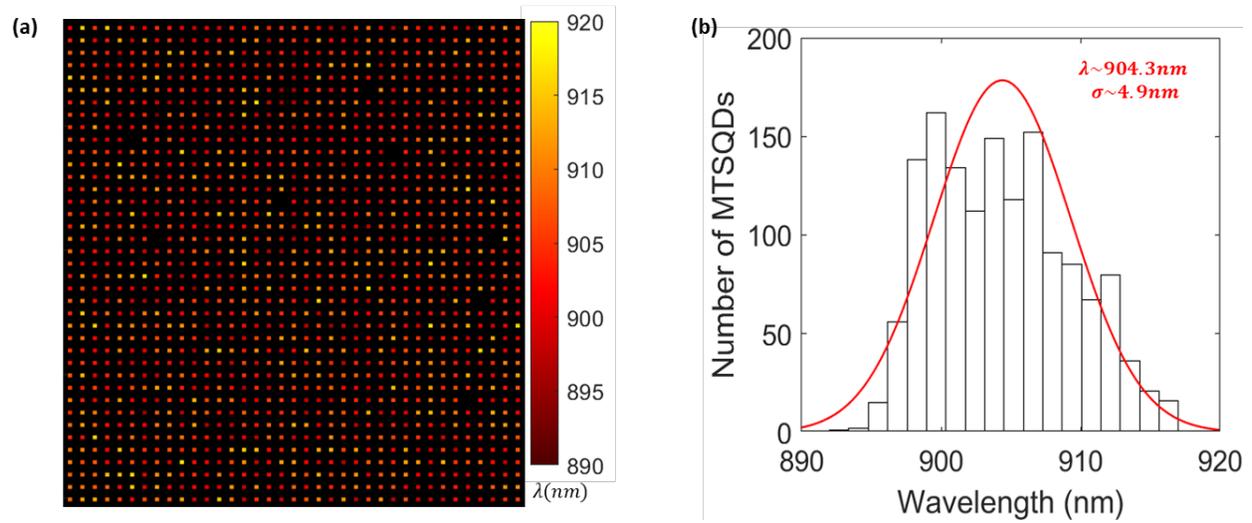

**Figure 2. Spectrally Uniform Scalable Arrays of MTSQDs.** **(a)** Color-coded image of the emission wavelengths from ~1400 MTSQDs in a 50x50 array (MTSQDs separated by a 5μm pitch). **(b)** Histogram of the emission wavelengths of (a) indicating mean emission (λ) at ~ 904.3nm with a ~ 4.9nm standard deviation (σ) as extracted from a Gaussian fit.

*2.3 Large Volume Shape-Controlled MTSQDs and Large Oscillator Strengths*

Time resolved photoluminescence (TRPL) characteristics of the neutral exciton in MTSQDs in (5x8) and (50x50) arrays, under resonant excitation, are shown in Figs. 3a and b, respectively. The TRPL results are fitted with a three-level system model accounting for fine-structure splitting [11] and thus allow us to extract the radiative decay lifetime of the MTSQD's neutral exciton. MTSQDs from both samples exhibit short decay lifetimes, with $T_1$=350ps for the MTSQD in (5x8) array and $T_1$=390ps for the MTSQD in (50x50) array. The oscillator strength of the QD's neutral exciton within the routinely invoked electric dipole approximation can be determined from its radiative decay lifetime using the formula: $f = \frac{6\pi\epsilon_0 m_0 c^3}{nT_1 F_p \omega^2 e^2}$ [30]. The measured $T_1$ from the MTSQDs in Fig. 3 results in $f$~29 for the MTSQD in the (5x8) array and $f$~27 for the MTSQD in the (50x50) array.

Most QDs in the literature, dominated by the lattice-mismatched strain-driven spontaneously formed 3D islands, SAQDs, exhibit a typical $T_1$~1ns and an oscillator strength of f~7-8. By contrast, the MTSQDs reported in Fig. 3 exhibit ~2.5 - 3 times shorter $T_1$ than SAQDs, indicating oscillator strengths 3-4 times larger than SAQDs [11, 31]. We hypothesize that the reproducible and scalable large oscillator strength of these MTSQDs comes from the ability to control the shape and large volume of MTSQDs, producing large confinement potential volumes as shown in Fig. 1e. To examine the attendant confinement potential depth, we further investigated the compositional characteristics of MTSQDs in both, the small (5x8) and large (50x50) arrays. The findings, shown in Figs. 4a and b, are discussed next.

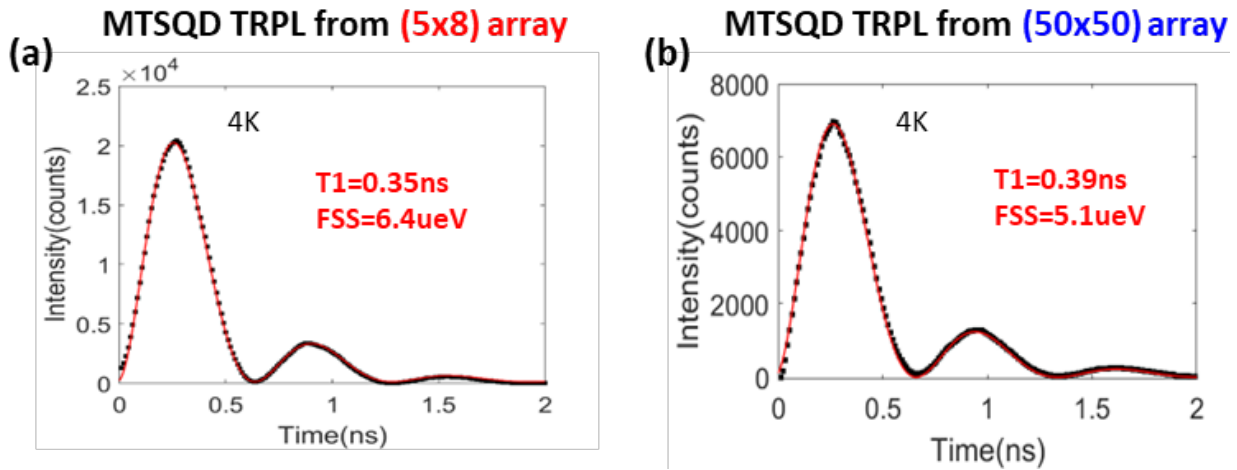

**Figure 3. Resonant TRPL from MTSQDs in (5x8) and (50x50) arrays.** **(a)** Resonant TRPL measurement of a large MTSQD from a (5x8) array showing fast decay lifetime ($T_1$=350ps) and large oscillator strength ($f$~29). **(b)** Resonant TRPL measurement of a large MTSQD from a (50x50) array (same array as in Fig. 2a) showing fast decay lifetime ($T_1$=390ps) and large oscillator strength ($f$~27).

Results from off-zone axis STEM-EDS mapping of In distribution of MTSQDs in both samples ((5x8) and (50x50) arrays) are shown in Figs. 4a and b. Both MTSQDs show large volume (with lateral sizes $\geq$ 30nm and heights $\geq$ 5nm) with uniform In distribution across the large-volume. A remarkable, yet not surprising, observation from the EDS data is the very similar shape of the In distribution in the MTSQDs from the different samples which comes from the ability to control the MTSQD shape with SESRE growth. In addition to this, we see that the In distribution is uniform across the large base of both MTSQDs which suggests a uniform in-plane confinement potential, a condition necessary for achieving weak-confinement of excitons in the MTSQD [4, 10]. Such results are the basis for the observed reproducible large oscillator strength in MTSQDs and the concomitant superradiant enhancement of light-matter interaction as discussed below.

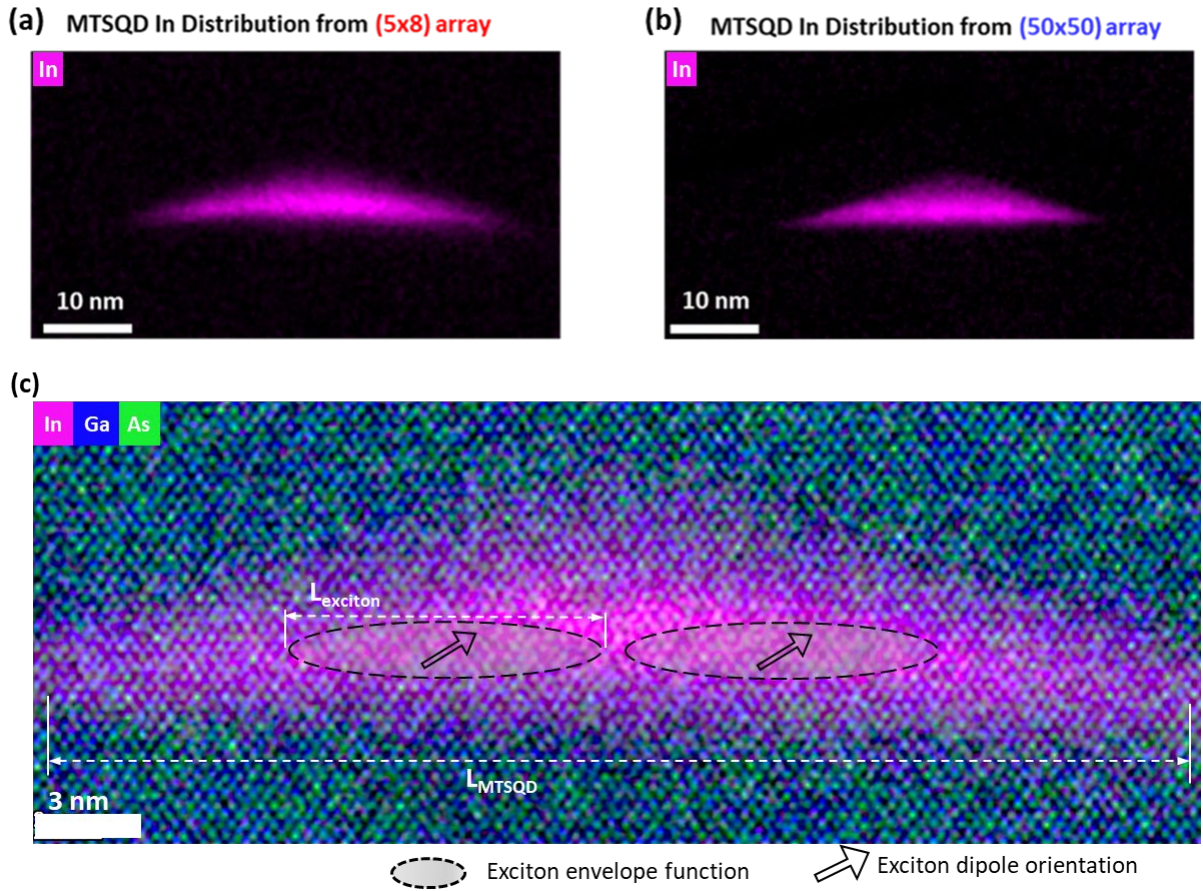

**Figure 4. Reproducible Large-Volume and Large Oscillator Strength of MTSQDs in (5x8) and (50x50) arrays.** **(a)** Off-zone axis EDS elemental map of a large MTSQD from a small (5x8) array, synthesized in the same growth but in a different array from the MTSQD in Fig. 2a. **(b)** Off-zone axis EDS elemental map of a large MTSQD from a large (50x50) array, synthesized in the same growth but in a different array from MTSQD in Fig. 2b. **(c)** Atomic-resolution EDS elemental mapping of In, Ga, and As distributions in the same MTSQD from Fig. 4a. From this, the height of the MTSQD is seen to be ~5nm, and the base length $L_{MTSQD}$ ~30nm. $L_{MTSQD}$ being larger than the exciton Bohr radius (~10nm) results in weak-confinement of excitons in MTSQDs ($L_{MTSQD} > L_{exciton}$) leading to superradiant enhancement of light-matter interaction, as discussed in the text.

*2.4 Single-Photon Superradiance in MTSQDs: Enhancement of Light-Matter Interaction*

In the concept of superradiance introduced by Dicke [32], a collective enhancement of light emission from an ensemble of N emitters occurs when the spatial distance between the emitters (in a given medium) is smaller than the wavelength of the electromagnetic field interacting with the emitters. In this regime, the emitters are coupled through the vacuum modes of the electromagnetic field resulting in a collective excitation of the emitters. The coherent phase relation between the emitters in the ensemble enhances the transition dipole since the excitation can now be localized in any individual emitter of the ensemble, resulting in the emission rate becoming proportional to N, the number of emitters in the ensemble. Although this concept was introduced in reference to an ensemble of emitters and has an extensive literature across a variety of physical systems, it is applicable at the level of a single photon and a single QD [10]. Briefly, in QDs, single photon emission arises from excitonic decay of a two-level system determined by the confinement volume of the single quantum dot. In single quantum dots the relevant length scales defining the enhancement of light emission originate in the exciton size ($L_{exciton}$) and the confinement potential size ($L_{QD}$), where the condition $L_{QD} > L_{exciton}$ (known as the weak-confinement regime in QDs) allows for the collective sharing of the exciton wavefunction across the N unit cells of the QD material, creating a giant transition dipole for the exciton and enhancing the attendant oscillator strength defining the light-matter interaction [9].

The behavior of the exciton in a QD, and thus the strength of light-matter interaction, can be dominated by either electron and hole state quantum-confinement energies in a deep confinement potential (regime of strong quantum confinement of one particle states) or, upon excitation, electron and hole Coulomb attraction effects which are determined by the size of the QD's confinement potential with respect to the exciton's Bohr radius [4, 8, 9]. In absence of the Coulomb force, the electron and hole within the QD move independently from each other, and can be characterized by respective envelope functions, $f_e(\bar{r}_e)$ and $f_h(\bar{r}_h)$. If the quantum confinement potential of electron and hole (determined by the size of the QD) is smaller than the Bohr exciton radius ($a_X$), the electron and hole motion remain decoupled, and this limit is referred to as strong exciton confinement. The result is that the exciton wavefunction becomes delocalized and spreads beyond the confinement volume, with the oscillator strength in the strong-confinement regime being then determined by the overlap of the electron and hole wavefunctions within the confinement volume (*i.e.* the QD volume), thereby limiting the maximum attainable oscillator strength to unity overlap between the electron and hole wavefunctions. By contrast when the QD size is larger than $a_X$, where the exciton size $L_{exciton}$ is roughly given by the Bohr exciton radius ($L_{exciton} = 2a_X$), the electron and hole motion is correlated by their mutual Coulomb attraction. In this regime, referred to as weak exciton confinement, the center of mass motion of the exciton becomes quantized and the relative motion between the electron and the hole can be assumed to be like the 1s hydrogen orbital, where the overall envelope function of the exciton can be expressed as [33],

$$F(\bar{r}_e, \bar{r}_h) = \frac{1}{\sqrt{N(a_X)}} e^{-\frac{|\bar{r}_e - \bar{r}_h|}{a_X}} f_e(\bar{r}_e) f_h(\bar{r}_h) \quad (2)$$

$$F_0(\bar{r}_e, \bar{r}_h) = f_e(\bar{r}_e) f_h(\bar{r}_h) \quad (3)$$

Critical is the additional factor $N(a_X)$ required to make sure that the exciton envelope function is normalized, and $N(a_X) < 1$ is indicative of the electron and the hole motion being correlated in space, resulting in an increased electron-hole overlap and enhanced light matter coupling strength. It has been shown that this increased overlap results in an enhancement of the transition dipole moment,[33] where the enhancement factor is $\frac{1}{\sqrt{N(a_X)}}$, and corresponds to the annihilation of the exciton, resulting in a factor of $\frac{1}{N(a_X)}$ enhancement in the oscillator strength.

As discussed throughout this paper, the superradiant enhancement of light-matter interaction in MTSQDs across the arrays comes from their large volume and the ability to control their shape and this is shown in the atomic-resolution EDS map of an MTSQD in Fig. 4c. The overlaid markings in Fig. 4c serve to capture the effect of the MTSQD's large-volume, and uniform three-dimensional confinement, on the exciton. In such weakly confining QDs, the motion of the electron and the hole becomes correlated and thus the exciton is able to sample a larger number of Bloch unit cells within the confinement volume resulting in a collective enhancement of light-matter interaction in the QD. This is illustrated in Fig. 4c with $L_{MTSQD} > L_{exciton}$ and the exciton envelope function being able to sample different regions of the MTSQD confinement potential at a time. Also shown in Fig. 4c is the fact that the coherent phase relation in the collective state shared by the exciton across the MTSQD volume comes from the exciton's dipole (with its orientation defined by the MTSQD's confinement potential), which leads to a giant transition dipole for the exciton and thus superradiant enhancement of light-matter interaction.

As a further validation of the enhancement of the oscillator strength owing to the superradiant effect in the MTSQDs, we employ the model shown in equation (2) and solve for the electron and hole envelope functions, under independent particle picture and under the effective mass approximation, for the MTSQD size and shape guided by the TEM studies. We construct a 3D confinement volume of pyramidal shape with elongated base diagonals (based on the known ~{103} bounding facets) [34], where we take the base size along [110] to be ~30nm, height to be ~5nm, and base size along [-110] to be ~10nm, and employ a finite element method calculation to obtain $f_e(\bar{r}_e)$ and $f_h(\bar{r}_h)$. We then include the electron-hole correlation ($e^{-\frac{|\bar{r}_e-\bar{r}_h|}{a_X}}$) with the Bohr exciton radius ~10 to 15nm (corresponding to Bohr exciton radius in bulk InGaAs) and numerically evaluate the oscillator strength enhancement factor $\frac{1}{N(a_X)}$ (For details, please see SI section S1). We find the oscillator strength enhancement to be ~2.5, consistent with the MTSQD oscillator strength (~30) being a factor of ~ 2.5 to 3 higher compared to the oscillator strength of typical InGaAs SAQDs that exhibit strong confinement. These results point to the uniqueness of the MTSQDs as a platform enabling control of light matter interaction by controlling the size and shape.

## 3 CONCLUSIONS

Beyond meeting all key figures-of-merit for technological advancement of photonic quantum information processing as a single photon source platform, both in the individual device level (brightness, purity, indistinguishability) and system level (scalability, spatial ordering, and spectral uniformity), MTSQDs are shown here to have controlled enhancement of light-matter interaction allowing for exploitation of superradiance phenomena. It is worth pointing out that MTSQDs are the only class of QDs that allows for tailoring the intrinsic oscillator strength and achieving reproducibly large $f$ in scalable and spatially-ordered arrays. Furthermore, the SESRE approach is implementable in a wide array of material systems, spanning from lattice-matched to -mismatched materials, allowing the benefits from precise positioning and enhanced light-matter interactions to be used in different emission wavelength regimes.

The essence of MTSQDs unique structural and optical properties lies in their synthesis through the SESRE growth approach. Results discussed in this paper highlight that the single photon superradiant enhancement of light-matter interaction observed in MTSQDs originates from the reproducible controlled large-volume and shape of MTSQDs in arrays, enabled by the control of run-to-run growth conditions and control of nanomesa profile evolution during SESRE growth. It is also important to emphasize that SESRE driven spatially-selective growth does not require lattice mismatch induced strain to achieve spatially-selective growth. Indeed, the first SESRE quantum dots were realized in the lattice-matched GaAs/AlGaAs material system [23].

## 4 EXPERIMENTAL METHODS

### 4.1 MTSQD Array Growth

The MTSQD samples studied in this paper were grown by molecular beam epitaxy (MBE) on GaAs (001) substrates patterned with pedestal nanomesas, arranged in arrays, sitting on top of a Distributed Bragg Reflector (DBR) mirror (as seen in Fig 1a). The samples, grown under the same growth conditions, differ only in the size of the nanomesa arrays used for growth; with one sample having small (5x8) arrays (40 MTSQDs per array) and the other sample having large (50x50) arrays (2500 MTSQDs per array). Each sample used for growth consisted of a substrate of size ~1cm x ~1cm containing different smaller areas (~1mm x 1mm) patterned with the nanomesa arrays. After growth the ~1cm x ~1cm substrate, now containing planarized MTSQD arrays, was cleaved into smaller pieces so that structural and optical characterization could be conducted on the different cleaved pieces from the same growth.

Before MTSQD growth, the GaAs (001) substrate containing the DBR structure[11] was patterned with arrays of square mesas of HSQ negative resist (~70nm thick), with HSQ mesa edges oriented along <100>, using electron-beam lithography (EBL, Raith EBPG 5150). The HSQ served as the mask for subsequent wet etching of the GaAs pedestal nanomesas. Etch rate of wet-etching solutions ($NH_4OH$ based) were carefully calibrated, allowing us to fabricate GaAs pedestal nanomesa arrays of with height ~100nm and lateral sizes ranging from ~100nm – ~400nm on both samples.

The growth structure (*i.e.* deposition layer sequence) and growth conditions were the same for both samples and consisted of: (i) deposition of 270ML of GaAs, at temperature of ~600C, As$_4$ flux of $P_{As4}$ ~ 2 x 10$^{-6}$ Torr, and growth rate of 0.25ML/s, which resulted in nanomesas developing a pyramidal morphology and an incumbent reduction of the mesa top size following SESRE (MTSQDs were targeted to form on nanomesas with starting lateral size of ~300nm, which after the 270ML GaAs deposition had a top size of ~30nm). (ii) deposition of 4.25ML of In$_{0.5}$Ga$_{0.5}$As, for MTSQD formation, at a temperature of ~520C, $P_{As4}$ ~ 3 x 10$^{-6}$ Torr, and growth rate of 0.5ML/s. Right after InGaAs deposition, 20ML of GaAs was deposited for pinching off the {103} sidewalls of the pyramidal nanomesas containing MTSQDs. (iii) deposition of ~1250ML of GaAs for planarizing the pyramidal nanomesa structures at ~600C and As$_4$ flux of $P_{As4}$ ~ 2 x 10$^{-6}$ Torr. Thin AlAs layers (~10ML) were deposited in regular intervals during the GaAs layer depositions to act as marker layers for STEM characterization as seen in Fig. 1a - d. Apart from the array size, the only difference between the samples studied here was that the number of AlAs marker layer deposited on the sample containing (50x50) arrays was doubled (with respect to the (5x8) array sample) for better inspection of the nanomesa profile evolution during STEM studies.

### 4.2 TEM Specimen Preparation

Preparation of specimens for STEM examination, carried out using a dual-beam focused ion beam (FIB) / scanning electron microscope (SEM) instrument (Helios G4 UXe PFIB), consisted of creating electron-transparent lamella containing the MTSQDs through a modified version of the typically employed site-specific lift-out technique. The main modifications in our approach compared to typical protocols were in (i) use of electron beam induced deposition (EBID) of Pt marker structures on the substrate surface (underneath which MTSQDs were located), before trench-milling and lift-out, allowing for controlled FIB thinning of the lifted-out lamella containing MTSQDs, and (ii) backscattered electron (BSE) imaging of the AlAs marker layers on the lamella surface during FIB thinning to control the specimen thickness and assure that the MTSQDs are not milled away during thinning (*i.e.* MTSQDs are contained within the lamella). This approach leads to reproducible control of the position of the MTSQD within the electron-transparent lamella every time TEM specimens are prepared. The TEM specimens prepared for the studies in this paper were all done along the [110] cross-section.

### 4.3 STEM and EDS Characterization

All STEM and EDS measurements presented here were conducted in a probe-corrected Spectra 200 X-CFEG STEM instrument. Atomic-resolution STEM imaging and EDS mapping conditions were chosen to maximize spatial resolution and thus were done with specimen oriented along the [1-10] zone axis and with a beam energy of 200keV, beam current ~100pA, beam semi convergence angle of 25mrad and probe aberrations corrected to $5^{th}$ order, resulting in a spatial resolution of <0.8 Å (as determined from the fast Fourier Transform (FFT) of the STEM images). STEM imaging was done using a high angle annular dark field (HAADF) detector set to inner and outer collection angles of ~80mrad – 200mrad for producing Z-contrast images arising from incoherent thermal diffuse scattering, minimizing diffraction-based contrast in the images. EDS data was acquired using a Dual-X detector, with a solid-angle of collection of ~1.8sr. Off-zone axis EDS measurements were done with a ~$4^0$ tilt away from the [1-10] zone axis (along the [200] direction), to reduce the effects of channeling, and with a beam current of 500pA to increase the counts from the x-ray characteristic peaks. The EDS elemental maps were calculated by extracting the integrated counts from x-ray characteristic peaks (*e.g.* In Lα1, Ga Kα1 and As Kα1) after removal of the Bremsstrahlung x-ray background counts.

**4.4 Optical Characterization**

Large area photoluminescence (PL) imaging was done on the (50x50) array (see Fig. 2a) using an in-house built tunable filter system. The measurement allows for imaging emission of individual MTSQDs in the array over large areas (with the are probed being limited by the excitation area of the beam, ~180um x 190um) and resolving their emission wavelength with a spectral resolution of ~1.6nm. Single photon measurements from individual MTSQDs (see TRPL data from Fig. 3) were done using a resonant excitation scheme, in vertical excitation and vertical detection geometry, with a spatial resolution ~1.2um and with the sample placed in a cryostat at 4K temperature. For details on the resonant excitation measurement setup and three-level model used for extracting radiative decay lifetimes from MTSQD neutral excitons please see ref. 11.

**Research Funding**

This work is supported by the Air Force Office of Scientific Research grant number FA9550-22 1-0376, the Center for Nanoimaging (CNI) at the University of Southern California, and the Kenneth T. Norris Professorship.

**Author Contributions**

All authors have accepted responsibility for the entire content of this manuscript and consented to its submission to the journal, reviewed all the results and approved the final version of the manuscript. LJ and QH grew the samples. LJ conducted the STEM and EDS characterization. QH, JZ and SL developed the optical instrumentation. QH conducted optical measurements. SC provided the theoretical analyses. AM coordinated and guided the overall project. All authors participated in manuscript writing.

**Conflict of Interest**

The authors declare no competing interests.

**Supporting Information Available:** Theoretical analysis of the enhancement of the oscillator strength of the exciton in MTSQDs.